\documentclass[fleqn,usenatbib]{mnras}

\usepackage{newtxtext,newtxmath}

\usepackage[T1]{fontenc}

\DeclareRobustCommand{\VAN}[3]{#2}
\let\VANthebibliography\thebibliography
\def\thebibliography{\DeclareRobustCommand{\VAN}[3]{##3}\VANthebibliography}

\usepackage{graphicx}	
\usepackage{amsmath}	

\DeclareMathOperator*{\argmax}{arg\,max}

\title[Vela small glitch]{A new small glitch in Vela discovered with a hidden Markov model}

\author[Dunn et al.]{
L. Dunn,$^{1,2}$\thanks{E-mail: liamd@student.unimelb.edu.au}
A. Melatos,$^{1,2}$
C. M. Espinoza,$^{3,4}$
D. Antonopoulou,$^{5}$ and
R. Dodson$^{6}$
\\
$^{1}$School of Physics, University of Melbourne, Parkville, VIC 3010, Australia\\
$^{2}$Australian Research Council Centre of Excellence for Gravitational Wave Discovery (OzGrav), University of Melbourne,\\
Parkville, VIC 3010, Australia\\
$^{3}$Departamento de F\'isica, Universidad de Santiago de Chile (USACH), Av. Victor Jara 3493, Estaci\'on Central, Chile\\
$^{4}$Center for Interdisciplinary Research in Astrophysics and Space Sciences (CIRAS), Universidad de Santiago de Chile, Chile\\
$^{5}$Jodrell Bank Centre for Astrophysics, School of Physics and Astronomy, The University of Manchester, Manchester M13 9PL, UK\\
$^{6}$International Centre for Radio Astronomy Research, University of Western Australia, Crawley, WA 6009, Australia
}

\pubyear{2023}

\begin{document}
\label{firstpage}
\pagerange{\pageref{firstpage}--\pageref{lastpage}}
\maketitle

\begin{abstract}
A striking feature of the Vela pulsar (PSR J0835$-$4510) is that it undergoes sudden increases in its spin frequency, known as glitches, with a fractional amplitude on the order of $10^{-6}$ approximately every 900 days.
Glitches of smaller magnitudes are also known to occur in Vela.
Their distribution in both time and amplitude is less well constrained but equally important for understanding the physical process underpinning these events.
In order to better understand these small glitches in Vela, an analysis of high-cadence observations from the Mount Pleasant Observatory is presented.
A hidden Markov model (HMM) is used to search for small, previously undetected glitches across 24 years of observations covering MJD 44929 to MJD 53647.
One previously unknown glitch is detected around MJD 48636 (Jan 15 1992), with fractional frequency jump $\Delta f/f = (8.19 \pm 0.04) \times 10^{-10}$ and frequency derivative jump $\Delta\dot{f}/\dot{f} = (2.98 \pm 0.01) \times 10^{-4}$.
Two previously reported small glitches are also confidently re-detected, and independent estimates of their parameters are reported.
Excluding these events, 90\% confidence frequentist upper limits on the sizes of missed glitches are also set, with a median upper limit of $\Delta f^{90\%}/f = 1.35 \times 10^{-9}$.
Upper limits of this kind are enabled by the semi-automated and computationally efficient nature of the HMM, and are crucial to informing studies which are sensitive to the lower end of the glitch size distribution.
\end{abstract}

\begin{keywords}
pulsars: individual: Vela -- stars: neutron -- stars: rotation
\end{keywords}



\section{Introduction}
The electromagnetic spindown of pulsars is sometimes interrupted by a glitch -- a sudden increase in the spin frequency.
At the time of writing, 670 glitches among 208 pulsars have been recorded in the Jodrell Bank Observatory (JBO) catalogue\footnote{\href{http://www.jb.man.ac.uk/pulsar/glitches/gTable.html}{http://www.jb.man.ac.uk/pulsar/glitches/gTable.html}} \citep{EspinozaLyne2011}.
Glitches provide a valuable window into the interior physics of neutron stars, but the underlying physical cause of glitches remains unknown; see \citet{HaskellMelatos2015} for a review of proposed mechanisms.

One of the most prolific glitching pulsars is the Vela pulsar (PSR J0835$-$4510), which has 24 recorded glitches since its discovery in 1968 \citep{LargeVaughan1968}.
Historically it has been a target of extensive monitoring [e.g. \citet{CordesDowns1988,DodsonLewis2007}] and many studies of glitching behaviour [e.g. \citet{DodsonMcCulloch2002,MelatosPeralta2008,ShannonLentati2016, PalfreymanDickey2018,EspinozaAntonopoulou2021}].
Vela has been monitored extensively by the Mount Pleasant Observatory near Hobart, Australia,  accumulating roughly two decades of high-cadence timing data between 1981 and 2005 \citep{DodsonLewis2007}.
This has proved to be a rich source of information on the timing behaviour of Vela, revealing several large glitches, including high time resolution observations of the large glitch of January 2000 \citep{DodsonMcCulloch2002}.
The high density of the observations also enables the identification of more subtle timing features.
Using later single-pulse observations of the 2016 Vela glitch from Mount Pleasant \citep{PalfreymanDickey2018}, \citet{AshtonLasky2019} related the observed, short-term, post-glitch recovery to the dynamics of three coupled components in the stellar interior.
Separately, \citet{EspinozaAntonopoulou2021} presented a systematic search for small glitches and other irregularities in Vela's rotation, finding two new small glitches as well as a population of small-amplitude changes in the spin frequency of both signs, which are regarded as a component of the timing noise \citep{CordesDowns1988, D'AlessandroMcCulloch1995, ChukwudeUrama2010}.

While timing data from Vela have been investigated extensively over the years, it is not common practice to employ Monte Carlo simulations to quantify the completeness of the glitch catalogues reported based on these data \citep{JanssenStappers2006, YuLiu2017}.
In this paper, we use a hidden Markov Model (HMM) to address this gap \citep{MelatosDunn2020}\footnote{\href{https://github.com/ldunn/glitch_hmm}{https://github.com/ldunn/glitch\_hmm}}.
The method has previously been successfully used to correct glitch parameter values whose estimation was confounded by periodic scheduling \citep{DunnLower2021}, and to set upper limits on the size of undetected glitches in datasets from timing programmes undertaken at both Parkes \citep{LowerJohnston2021} and UTMOST \citep{DunnMelatos2022}.

We provide an overview of the HMM glitch detector in Section \ref{sec:hmm}.
Upon conducting an automated HMM search for missed glitches between Nov 21 1981 and Oct 24 2005, we discover one small new glitch, discussed in Section \ref{sec:small_glitch}.
In Section \ref{sec:other_small_gl} we present re-detections and re-analyses of the two small glitches previously detected by \citet{EspinozaAntonopoulou2021}.
Finally, we put systematic frequentist upper limits on the size of any further missed glitches in our Vela dataset, as discussed in Section \ref{sec:undetected}.

\section{HMM glitch detector}
\label{sec:hmm}
In this section we give a brief outline of the HMM glitch detection algorithm.
The reader is referred to \citet{MelatosDunn2020} for a complete description, and to \citet{Rabiner1989} for an overview of HMMs.

\subsection{Pulsar timing in an HMM framework}
\label{subsec:hmm_basics}
HMMs assume that the system in question has a well-defined internal state at each timestep, which evolves according to some Markov process.
However, the internal state is ``hidden''; we rely on some  proxy measurement which implies a probability of occupying each hidden state.
In this paper the system is the Vela pulsar, and the internal state is the combination of its spin frequency $f$ and frequency derivative $\dot{f}$ (discretised on a grid).
The measurements are the barycentred pulse times of arrival (ToAs), which are connected to the underlying hidden states by the requirement that the number of rotations accumulated between two consecutive ToAs should be an integer (allowing for small deviations away from this requirement due to uncertainty on the ToAs and the discretisation of $f$ and $\dot{f}$).
This requirement is represented in the HMM through the ``emission probability'', which expresses the probability of observing a particular measurement (namely the gap between consecutive ToAs, denoted by $z$) given the system is in a particular hidden state ($f$ and $\dot{f}$ pair, denoted by $q_i$).
Mathematically, we write the emission probability $\Pr(z \mid q_i)$ in terms of a von Mises distribution \begin{equation} \Pr(z \mid q_i) = \frac{\exp[\kappa\cos(2\pi\Phi)]}{2\pi I_0(\kappa)}, \label{eqn:hmm_emission} \end{equation} where $\Phi$ denotes the number of turns accumulated during the ToA gap $z$ given an $(f, \dot{f})$ pair, $\kappa$ is a parameter containing the uncertainty in the phase due to both ToA uncertainty and the discretisation of $f$ and $\dot{f}$, and $I_0(x)$ is the zeroth modified Bessel function of the first kind.
Explicit expressions for $\Phi$ and $\kappa$ as a function of $q_i$ are given in equations (1) and (2) of \citet{DunnMelatos2022}, and the mathematical form of $\Pr(z\mid q_i)$ given in equation (\ref{eqn:hmm_emission}) is justified in \citet{MelatosDunn2020}.

The Markov process which drives the evolution of the hidden state is assumed in this work to be a random walk in the frequency derivative, i.e. \begin{equation} \frac{\mathrm{d}^2f}{\mathrm{d}t^2} = \xi(t), \label{eqn:hmm_f_evo} \end{equation} where $\xi(t)$ is a white noise term satisfying \begin{align} \langle \xi(t) \rangle &= 0, \\\langle \xi(t)\xi(t')\rangle &= \sigma_{\text{TN}}^2\delta(t-t'), \label{eqn:hmm_sigma_tn}\end{align} where $\sigma_{\text{TN}}$ is a free parameter which determines the strength of the timing noise included in the model.
Equation (\ref{eqn:hmm_f_evo}) applies in the interval between glitches.
At any instant where a glitch occurs, $f$ and $\dot{f}$ jump discontinuously by $\Delta f_{\mathrm{p}}$ and $\Delta\dot{f}_{\mathrm{p}}$ respectively (where p denotes permanent; see Section \ref{subsec:hmm_glitch_measurement}).
The no-glitch ephemeris generated by traditional timing methods provides an approximation to the initial conditions satisfied by equation (\ref{eqn:hmm_f_evo}).

In the above approach to pulsar timing we track the underlying spin evolution based on measurements of the phase locally, as inferred from consecutive ToAs, rather than construct a model of the phase globally.
Approaching this task using an HMM allows us to take advantage of existing algorithms to efficiently compute useful quantities such as the posterior distributions on $f$ and $\dot{f}$, as well as the probability of an observed data set, given a particular model (i.e. the Bayesian evidence).
The computational framework provided by the HMM formalism is a central motivation for adopting this approach.

\subsection{Glitch detection with the HMM}
\label{subsec:hmm_glitch_det}
As mentioned in the previous section, the structure of the HMM makes it computationally efficient to compute the model evidence $\Pr(D \mid M)$ where $D$ is the data analysed and $M$ is some particular setup of the HMM.
The algorithm used to perform this calculation is known as the forward algorithm \citep{Rabiner1989, MelatosDunn2020}.
In particular, we can specify two classes of models.
In one, denoted $M_0$, the evolution of the pulsar's rotation is governed solely by a combination of secular spindown and timing noise.
In the other, denoted $M_n(k_1, k_2, \ldots, k_n)$, the evolution of the pulsar's rotation is governed by the secular spindown and timing noise \emph{except} during $n$ ToA gaps containing glitches indexed by $\{k_1, k_2, \ldots, k_n\}$.
The characteristics of the glitch are not specified by the model except that the frequency jump $\Delta f_{\mathrm{p}}$ is assumed to be instantaneous and constrained to be positive.
An instantaneous jump in frequency derivative $\Delta\dot{f}_{\mathrm{p}}$ is also allowed, and may be of either sign.

With the above setup, model selection proceeds by computing the Bayes factors \begin{equation} K_1(k) = \frac{\Pr[D \mid M_1(k)]}{\Pr(D \mid M_0)}. \end{equation}
These $K_1(k)$ are compared against a detection threshold $K_\text{th}$ which is chosen ahead of time according to the analyst's preference.
Here we follow \citet{MelatosDunn2020} and \citet{DunnMelatos2022} in choosing $\ln(K_\text{th}) = 1.15$ (corresponding to $K_\text{th} = 10^{1/2}$).
The possibility of multiple glitches in a dataset is handled via the greedy procedure described in section 4.2 of \citet{MelatosDunn2020}.
Briefly, if at least one $K_1(k)$ exceeds $K_\text{th}$, we choose \begin{equation}k^* = \argmax_{2 \leq k \leq N_T-1} K_1(k)\end{equation} and proceed to calculate the Bayes factors \begin{equation} K_2(k^*, k_2) = \frac{\Pr\left[D \mid M_2(k^*, k_2)\right]}{\Pr\left[D \mid K_1(k^*)\right]}\end{equation} for $2 \leq k_2 \neq k^* \leq N_T -1$.
This procedure repeats until none of the computed Bayes factors exceed $K_\text{th}$.

\subsection{Glitch measurement with the HMM}
\label{subsec:hmm_glitch_measurement}
\begin{figure}
    \centering
    \includegraphics[width=\columnwidth]{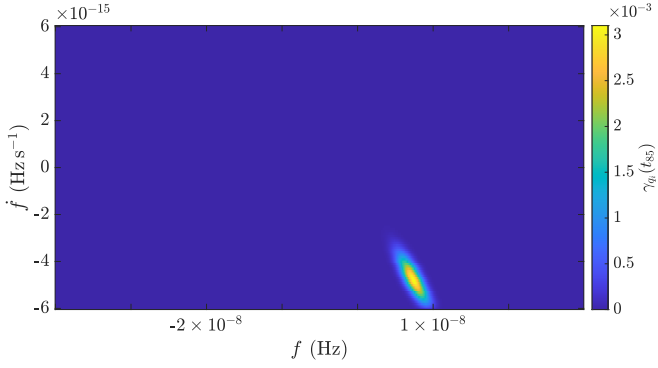}
    \caption{Illustrative heatmap showing the hidden state posterior $\gamma_{q_i}(t_n)$ for a single timestep in the analysis of the glitch event described in Section \ref{sec:small_glitch}. The color scale reflects the posterior density at the 85th timestep, $\gamma_{q_i}(t_{85})$. The plotted range in $f$ and $\dot{f}$ reflects the complete allowed space of hidden $(f, \dot{f})$ states in this analysis, and the hidden state posterior is well-localised within this range.}
    \label{fig:full_post_heatmap}
\end{figure}

In order to measure the glitch properties using the HMM, we require estimates of the value of the hidden state as a function of time.
To do this, we use the forward-backward algorithm \citep{Rabiner1989, MelatosDunn2020}, an extension of the forward algorithm which is used to calculate the model evidences, to obtain the posterior distribution on the hidden state $q_i = (f, \dot{f})$ at each timestep.
This distribution is denoted by $\gamma_{q_i}(t_n)$.
The $t_n$ are the timestamps associated with the observed data used in this analysis: the barycentred times of arrival.
Thus we emphasise that $\gamma_{q_i}(t_n)$ is evaluated at a discrete set of times typically separated by one to a few days.
An exemplar heatmap of $\gamma_{q_i}(t_n)$ is shown in Figure \ref{fig:full_post_heatmap}.
From this posterior we construct the one-dimensional posteriors for $f$ and $\dot{f}$ by marginalising over $\dot{f}$ and $f$, respectively denoted $\gamma_f(t_n)$ and $\gamma_{\dot{f}}(t_n)$.

We take the mode of $\gamma_f(t_n)$ at each timestep to construct the sequence of most likely frequencies during each ToA gap, denoted $f^*(t_n)$.
An analogous procedure is used to construct the sequence of most likely frequency derivatives, $\dot{f}^*(t_n)$.
For the purposes of parameter estimation, our model for the phase induced by a glitch includes a permanent (indicated by a subscript p) step change in phase\footnote{The step change in phase in equation (\ref{eqn:gl_with_exp}) does not represent a physical process.
Rather it allows us to account for an incorrect estimate of the glitch epoch $t_\text{g}$ in a manner which is relatively easy to fit. A physical step change in phase may occur if, for example, there is a change in the magnetic longitude of the pulsar's radio beam due to magnetospheric processes at the time of the glitch, e.g. \citet{PalfreymanDickey2018}, but we do not consider such processes here.}, frequency, and frequency derivative at the epoch of the glitch $t_{\mathrm{g}}$, as well as an exponentially decaying component (indicated by the subscript d): 
\begin{equation}
    \begin{split}
        \Delta\phi_\text{g}(t) = \Theta(t-t_\text{g})\{&\Delta\phi + \Delta f_\text{p}(t-t_\text{g}) + \frac{1}{2}\Delta\dot{f}_\text{p}(t-t_\text{g})^2 \\
        &  \left. + \tau_\text{d}\Delta f_\text{d}\left[1 - e^{-(t-t_\text{g})/\tau_\text{d}}\right]\vphantom{\int}\right\}.
    \end{split} 
    \label{eqn:gl_with_exp} 
\end{equation}
In equation (\ref{eqn:gl_with_exp}), $\Theta(x)$ denotes the Heaviside step function.
With the exception of the glitch described in Section \ref{subsec:1999_gl}, we set $\Delta f_\text{d} = 0$.
To estimate the glitch parameters from the HMM output we fit the time derivative of equation (\ref{eqn:gl_with_exp}) to the post-glitch values of $f^*(t_n)$ using the \textsc{lmfit} package \citep{NewvilleStensitzki2014}.
The $f^*(t_n)$ values are assigned uncertainties based on the width of $\gamma_f(t_n)$, estimated as the standard deviation of a Gaussian distribution fitted to $\gamma_f(t_n)$.
These Gaussian fits are also performed using \textsc{lmfit}.

The HMM differs from the traditional method of fitting a global phase model to the ToAs.
Although the HMM incorporates phase information from the ToAs directly through $\Pr(z \mid q_i)$ (see Section \ref{subsec:hmm_basics}), it returns information about the evolution of $f$ and $\dot{f}$, rather than parameters of a phase model which applies over the entire dataset.
Hence if we are to estimate the parameters of the glitch using the HMM we are obliged to model the post-glitch frequency evolution rather than phase evolution.
Glitch parameter estimation based on the values of $f^*(t_n)$ and $\dot{f}^*(t_n)$ has been performed in previous analyses \citep{DunnLower2021, DunnMelatos2022}.
The results therein as well as in Sections \ref{sec:small_glitch} and \ref{sec:other_small_gl} of this work return glitch parameters broadly in line with those returned by traditional \textsc{tempo2} or \textsc{temponest}-based analyses.
In all cases the parameter estimates should be understood as conditional on the model assumed.
In the case of the HMM this includes both the HMM's model of the dynamics of $f$ and $\dot{f}$ [e.g. equation (\ref{eqn:hmm_f_evo})] as well as the model of post-glitch frequency evolution used [viz. equation (\ref{eqn:gl_with_exp})].

\section{A small glitch at MJD 48636}
\label{sec:small_glitch}
Here we describe the detection and subsequent analysis of a new small glitch in the Vela pulsar.
The data analysed are described in sections 2 and 3 of \citet{EspinozaAntonopoulou2021}.
In particular we analyse the daily effective times of arrival described in section 3.2, rather than the full dataset which contains hundreds to thousands of ToAs per day.
Using the full dataset would inflate the computational cost but would not be expected to enhance the sensitivity of the HMM glitch detector; see Section \ref{subsec:cadence} for a brief discussion of the effect of observing cadence on sensitivity.
In total the data cover MJD 44929 to MJD 53667, and are broken down into smaller sections, typically using either known glitches or large observing gaps as boundaries.
Where possible, the data sections are allowed to overlap somewhat to mitigate the chance of missing a glitch which lies close to a section boundary.

For each section we use \textsc{tempo2} to fit values of $f$, $\dot{f}$ and $\ddot{f}$ to incorporate as an initial condition the local, secular evolution of the pulsar in the HMM analysis.
In a few cases where the data section begins immediately after a large glitch and its associated recovery, we exclude the first ${\sim} 50\,\mathrm{days}$ of data to avoid the majority of the recovery, which the HMM is unable to handle without artificially inflating the strength of the timing noise tracked by the HMM [see equation (\ref{eqn:freq_wandering_tn})].
\citet{ShannonLentati2016} found evidence for a common recovery timescale in PSR J0835$-$4510 of $25\,\mathrm{days}$ (in addition to a much longer timescale of $1300\,\mathrm{days}$ which the HMM is not confounded by), so by excluding $50\,\mathrm{days}$ we expect to exclude approximately $86\%$ of the recovery (i.e. two $e$-foldings).
Details of the data sectioning are given in Table \ref{tbl:sections}.

\begin{table}
    \centering
    \begin{tabular}{lrr}\hline
        Section & MJD start & MJD end\\\hline
        1 & 44964 & 45191\\\hline
        2 & 45211 & 45446\\
        3 & 45421 & 46257\\\hline
        4 & 46285 & 46562\\\hline
        5 & 47894 & 48049\\
        6 & 48030 & 48456\\\hline
        7 & 48475 & 48549 \\\hline
        8* & 48551 & 48719 \\
        9 & 48680 & 49558 \\\hline
        10 & 49560 & 49590\\\hline
        11 & 49594 & 49919 \\
        12 & 49882 & 50367 \\\hline
        13 & 50370 & 50703 \\
        14 & 51294 & 51424 \\\hline
        15 & 51425 & 51558 \\\hline
        16 & 51559 & 52039 \\
        17 & 52001 & 52339\\
        18 & 53095 & 53192 \\\hline
        19 & 53194 & 53218 \\
        20 & 53223 & 53500 \\
        21 & 53528 & 53667\\\hline
    \end{tabular}
    \caption{Details of the data sectioning used for the HMM glitch search. Horizontal rules indicate the location of known glitches. The asterisk indicates the data section containing the newly discovered glitch described in Section \ref{sec:small_glitch}.
    \label{tbl:sections}}
\end{table}
\subsection{Detection of the new glitch}
As discussed in Section \ref{subsec:hmm_glitch_det}, the HMM detects glitches by Bayesian model selection, comparing models which contain a glitch in a given ToA gap to the model which contains no glitches.

\begin{figure}
    \centering
    \includegraphics[width=0.9\columnwidth]{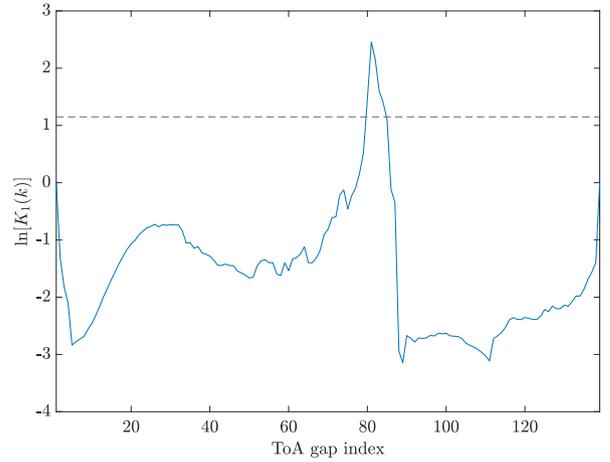}
    \caption{Sequence of $\ln\left[K_1(k)\right]$ for each ToA gap in the section of data containing the glitch described in Section \ref{sec:small_glitch}. A clear peak is visible at the 81st ToA gap, corresponding to MJD $48635.6$ -- MJD $48636.6$. The threshold $\ln(K_\text{th}) = 1.15$ is shown as the dashed line.}
    \label{fig:small_glitch_bfs}
\end{figure}

In analysing the stretch of data between MJD $48551$ and MJD $48719$, the HMM returns a glitch candidate between the ToAs at MJD $48635.6$ and MJD $48636.6$, with a log Bayes factor $\ln[K_1(k^*)] = 2.4$ ($k^* = 81$).
Fig. \ref{fig:small_glitch_bfs} shows $\ln[K_1(k)]$ for each ToA gap, indexed by $k$. A clear peak is visible at $k^* = 81$.
The HMM is sensitive to even single ToAs which are significantly displaced from the trend.
Therefore we check that the glitch candidate is not due to a non-astrophysical disturbance of the surrounding ToAs by re-running the analysis with the two ToAs bracketing the glitch candidate removed. 
A log Bayes factor of $2.9$ is returned, confirming that the candidate is not simply due to a disturbed ToA in the immediate vicinity of the glitch.
The veto procedure of removing the bracketing ToAs is necessary because the HMM cannot itself distinguish between a ToA which is in error due to radio frequency interference (RFI), or an observation with low signal-to-noise ratio, and so on, and a ToA which is displaced because of a glitch.
The HMM treats all ToAs equally.
In accommodating a displaced ToA it may be that the HMM inference prefers a glitch-containing model which ``explains'' the displaced ToA via some change in $(f, \dot{f})$ over the model with no glitch, and hence a spurious glitch candidate is returned.
Furthermore, due to the inclusion of spin-wandering in the HMM via equation (\ref{eqn:hmm_f_evo}), an incorrect model which includes a spurious change in $(f, \dot{f})$ is free to wander back to the true spin evolution over the longer term.
Hence the inclusion of data taken long after the RFI event does not significantly penalise the incorrect model.
This is in contrast to a model which does not include spin wandering, where a change in $(f, \dot{f})$ induces phase residuals which grow monotonically over time.
We expect $t_\text{g}$ to be either immediately before or after a displaced ToA, so re-running the search with the ToAs bracketing the glitch candidate removed eliminates the RFI scenario.

We also check that dispersion measure (DM) effects are not masquerading as a glitch.
We measure the evolution of the DM over the data span containing the glitch candidate using the high-cadence ToAs at $635$ and $950\,\mathrm{MHz}$ [with a cadence of approximately 2 minutes at both observing frequencies covering about 70\% of each day \citep{EspinozaAntonopoulou2021}], looking for sharp changes in the DM around the epoch of the glitch candidate.
No such changes are apparent and we find that a linear model with a DM of $68.288\,\mathrm{pc}\,\mathrm{cm}^{-3}$ at MJD $48616$ and a slope of $-0.036\,\mathrm{pc}\,\mathrm{cm}^{-3}\,\mathrm{yr}^{-1}$ describes the data well. 
This is a minor change from the original DM model which is constant at $68.29\,\mathrm{pc}\,\mathrm{cm}^{-3}$ for the section of data containing the glitch candidate.
Re-analysing the data with the refined, linearly-sloping DM model increases the log Bayes factor of the candidate marginally to $2.5$, and we use the refined DM model in the rest of the analysis.
Finally, we check for further glitches in this stretch of data following the greedy procedure outlined in Section \ref{subsec:hmm_glitch_det}.
None of the $K_2(k^*, k_2)$ exceed $K_\text{th}$.

The next most significant glitch candidate over the full 24 years of data has $\ln[K_1(k)] = 0.3$. 
The mean value of $\ln[K_1(k)]$ across the full 24 years (excluding the stretch of data containing MJD 48636) is $\langle\ln[K_1(k)]\rangle = -2.21$, with a standard deviation of $1.05$.
The glitch candidate at MJD 48636 is thus a clear outlier, and unlikely to simply be a timing noise excursion.
The event at MJD 48636 was also flagged in the analysis performed by \citet{EspinozaAntonopoulou2021}, but was ultimately discarded.
Its characteristics, and the grounds on which it was discarded, are summarised in Appendix \ref{apdx:espinoza_48636}.

\subsection{Parameter estimation}
\label{subsec:new_gl_params}
\begin{figure*}
    \centering
    \includegraphics[width=0.45\linewidth]{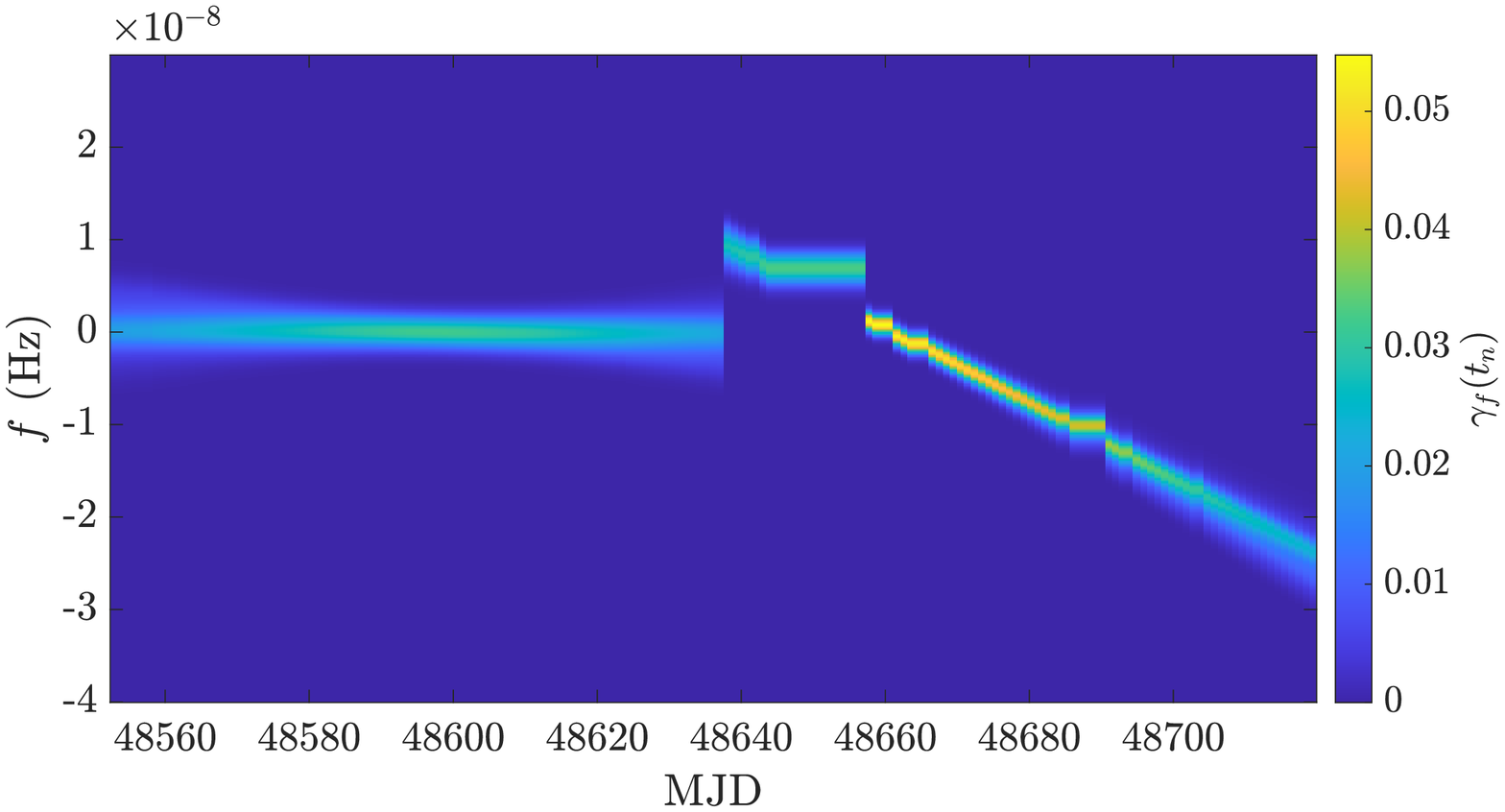} \includegraphics[width=0.45\linewidth]{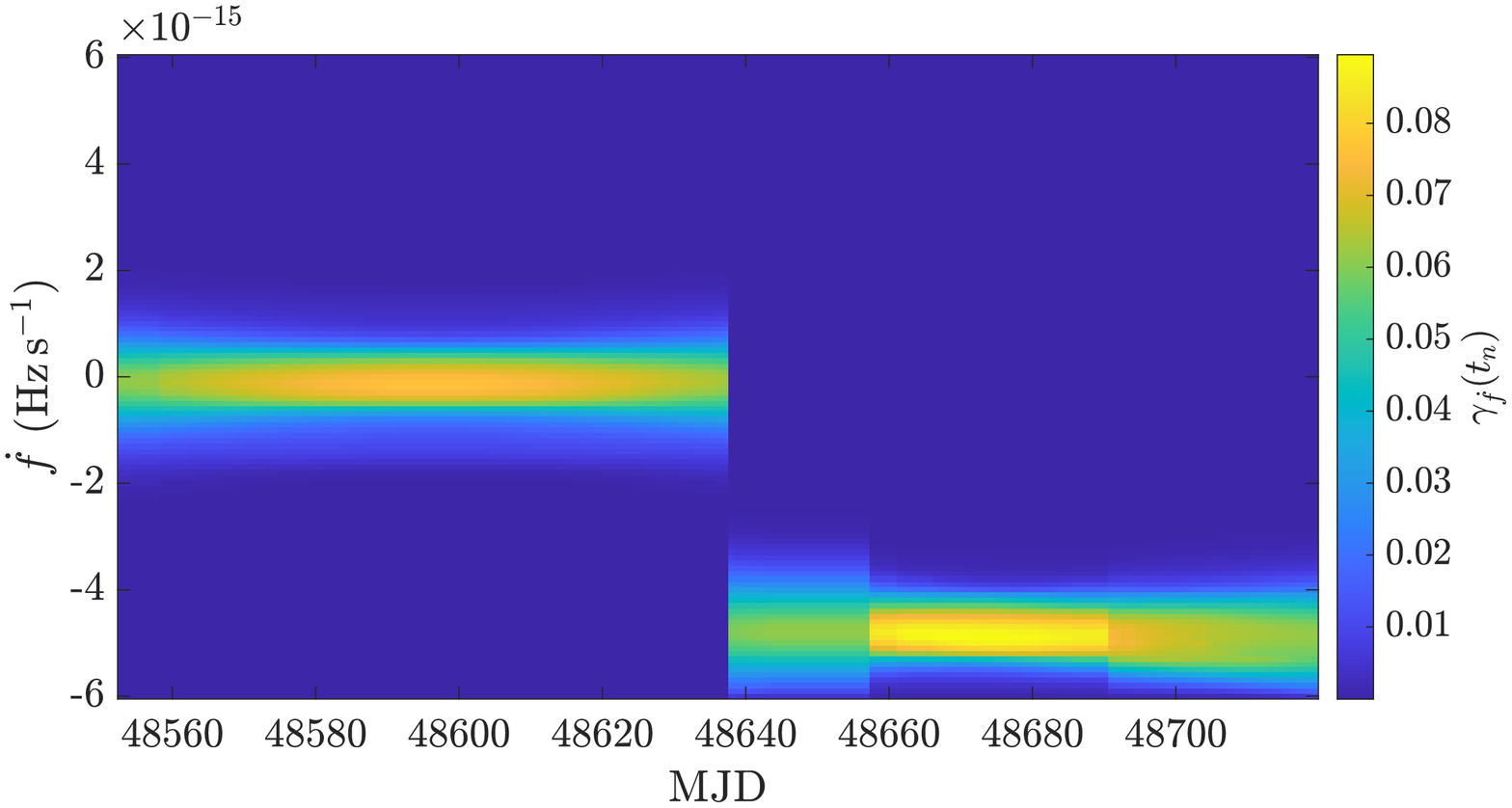}\\
    \includegraphics[width=0.45\linewidth]{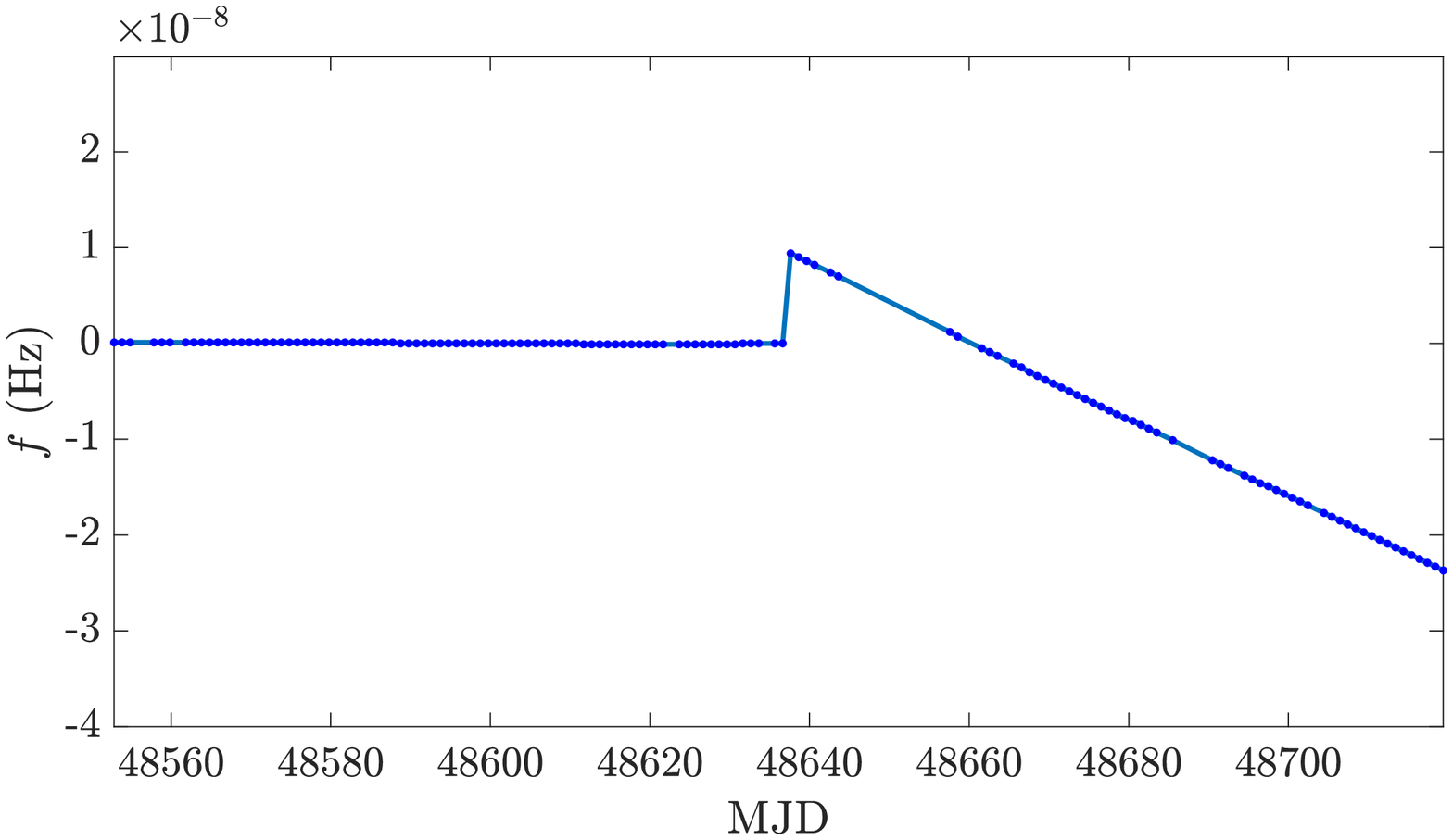} \includegraphics[width=0.45\linewidth]{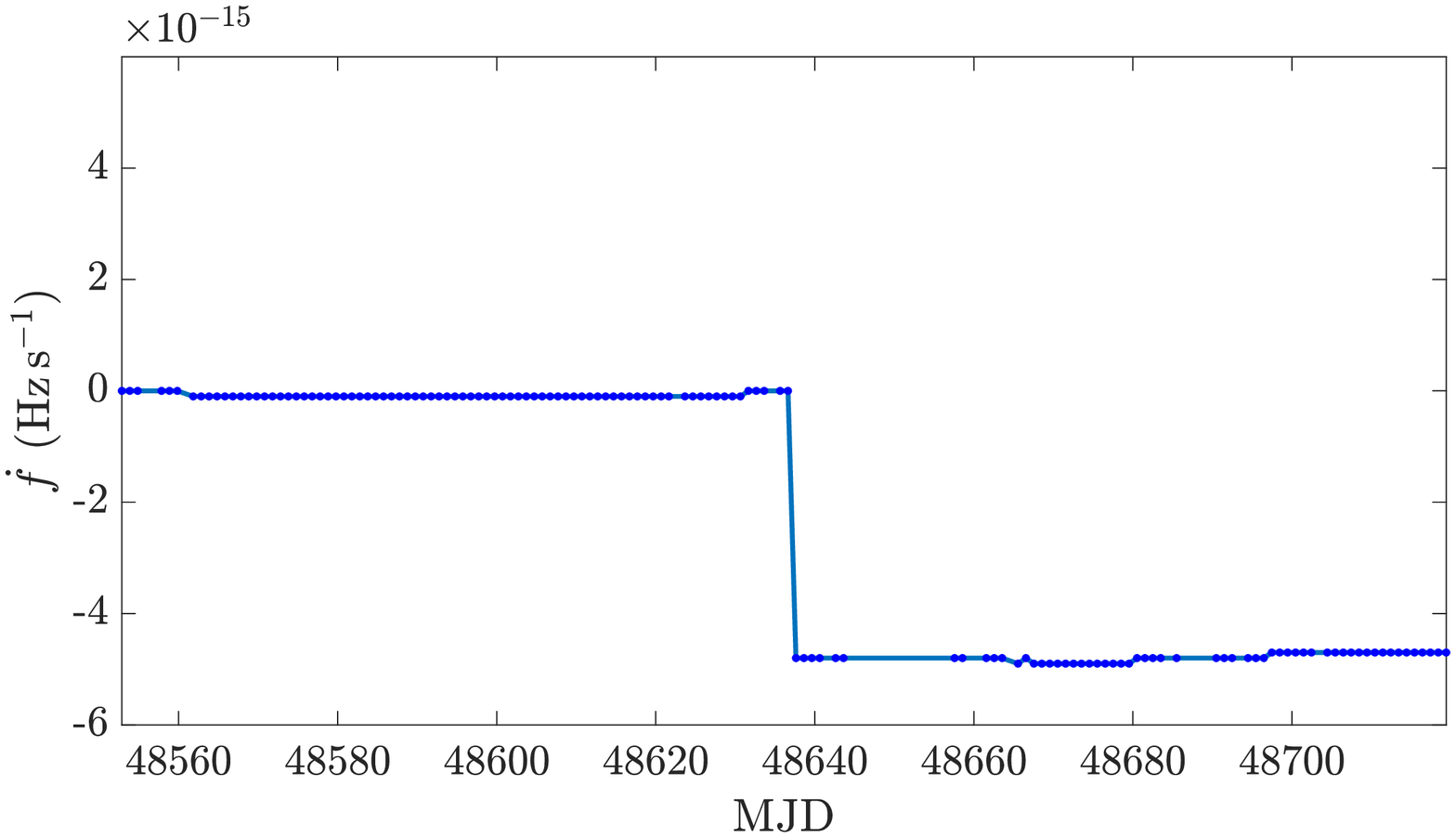}
    \caption{\emph{(Top)} Heatmaps of the marginalised posterior distributions of $f$ \emph{(left)} and $\dot{f}$ \emph{(right)} for the data section containing the newly discovered glitch described in Section \ref{subsec:new_gl_params}. \emph{(Bottom)} Sequence of \emph{a posteriori} most probable $f$ \emph{(left)} and $\dot{f}$ \emph{(right)} states over time, for the same data section. The values of $f$ and $\dot{f}$ are relative to the pre-glitch timing solution, hence they are flat and centred on zero in the pre-glitch region.}
    \label{fig:48636_gl_results}
\end{figure*}

With the detection of the new glitch established, we move on to estimate its parameters, following the procedure outlined in Section \ref{subsec:hmm_glitch_measurement}.
The key outputs from the HMM for this purpose are the marginalised posteriors of $f$ and $\dot{f}$, heatmaps of which are shown in the top two panels of Fig. \ref{fig:48636_gl_results}.
We note two features of these heatmaps, which also apply to the analogous figures for the glitch events described in Section \ref{sec:other_small_gl}, namely Figs. \ref{fig:1991_gl_results} and \ref{fig:1999_gl_results}.
First, the value shown on the heatmaps between $t_n$ and $t_{n+1}$ is the value of the posterior distribution evaluated at $t_n$.
For long ToA gaps this can give the impression that the frequency remains constant for many days in certain parts of the post-glitch heatmap, for example between MJD 48650 and MJD 48660 in the top left panel of Fig. \ref{fig:48636_gl_results}.
We prefer for transparency to have this aspect of $\gamma_{q_i}(t_n)$ reflected in the heatmaps, rather than choose an interpolation scheme which artificially evolves $\gamma_{q_i}(t_n)$ over a single ToA gap to give the impression of smooth $f$ and $\dot{f}$ evolution.
Second, the posteriors visibly widen at either end of the data span, and close to the epoch of the glitch $t_\text{g}$.
This is a real effect, because at those times the evolution of the hidden state is less constrained.
Typically the posterior at a given time $t_n$ is constrained by the data preceding and following $t_n$, and data which are further separated from $t_n$ are less constraining.
However, the posterior at $t_n < t_\text{g}$ is not constrained by the data at $t_n > t_\text{g}$, because of the considerable freedom in both $f$ and $\dot{f}$ evolution allowed at $t_\text{g}$. 
The same argument applies to posteriors calculated at $t_n > t_\text{g}$, which are not constrained by the data at $t_n < t_\text{g}$.

We also note the small fluctuations in $\dot{f}^*(t_n)$ which are visible in the bottom right panel of Fig. \ref{fig:48636_gl_results}.
These are real in the sense that they are not visual artefacts, and correspond to genuine changes in the mode of $\gamma_{\dot{f}}(t_n)$ by one $\dot{f}$ bin at a time.
However, given that the scale of the fluctuations is much smaller than the width of $\gamma_{\dot{f}}(t_n)$, we do not ascribe any physical significance to these features.

Parameter estimation with the HMM proceeds by fitting the maximum \emph{a posteriori} values of the spin frequency, $f^*(t_n)$, against the model for the post-glitch frequency evolution given by the time derivative of equation (\ref{eqn:gl_with_exp}).
Note that when performing the latter analysis, the values of $f$, $\dot{f}$ and $\ddot{f}$ are fitted using only the pre-glitch data.
This is in contrast to the HMM analyses searching for undetected glitches where the values are derived using the complete data in each section (at this stage, there is no notion of pre- and post-glitch, as a glitch has not yet been detected).
Fitting to the pre-glitch data ensures that the derived $f^*(t_n)$ and $\dot{f}^*(t_n)$ tracks are approximately flat before the glitch occurs, which simplifies the parameter estimation.

As a control experiment, we also estimate the glitch parameters using \textsc{temponest} \citep{LentatiAlexander2014}, which includes both deterministic terms [$f$, $\dot{f}$, $\ddot{f}$, and the glitch parameters $\Delta\phi$, $\Delta f_{\mathrm{p}}$, and $\Delta\dot{f}_{\mathrm{p}}$ in equation (\ref{eqn:gl_with_exp})] and a model of the phase residuals due to timing noise as a red noise process with power spectral density \begin{equation} P(f) = \frac{A^2}{12\pi^2}\left(\frac{f}{f_\text{yr}}\right)^{-\beta}, \label{eqn:tn_rn_psd}\end{equation} where $f_{\mathrm{yr}} = (1\,\mathrm{yr})^{-1}$.

\begin{figure}
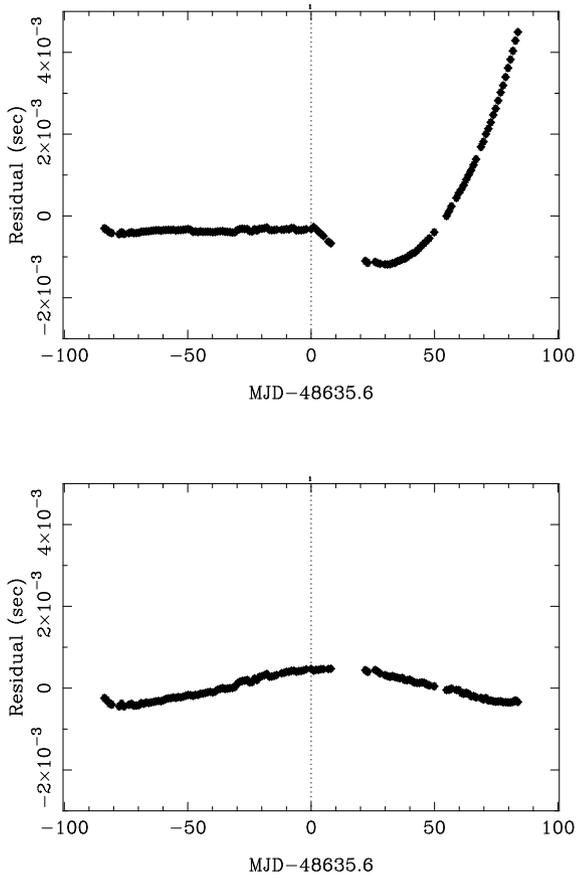

    \centering
    \includegraphics[width=0.9\columnwidth]{figures/tempo2_noglitch_fixedy.eps}\\
    \includegraphics[width=0.9\columnwidth]{figures/tempo2_glitch_fixedy.eps}
    \caption{Phase residuals from \textsc{tempo2} before \emph{(top)} and after \emph{(bottom)} fitting for the new small glitch at MJD $48636$ with \textsc{temponest}. The location of the glitch is marked by the vertical dotted line. Note that the post-fit residuals are not flat, because the \textsc{temponest} fit includes a timing noise component which is not subtracted from the residuals.}
    \label{fig:tempo_residuals}
\end{figure}

The results of the HMM and \textsc{temponest} fits are reported in Table \ref{tbl:48636_gl_params}.
\begin{table}
    \centering
    \begin{tabular}{lrr}\hline
        Parameter (units) & HMM fit (via \textsc{lmfit}) & \textsc{temponest} fit \\\hline
        $t_\text{g}$ (MJD) & 48635.64 &  48635.64 \\
        $\Delta\phi$ (turns) & N/A & $(-1.5 \pm 0.8) \times 10^{-3}$ \\
        $\Delta f_\text{p}$ (Hz) & $(9.24 \pm 0.03) \times 10^{-9}$ & $(9.3 \pm 0.8) \times 10^{-9}$ \\
        $\Delta\dot{f}_\text{p}$ ($\mathrm{Hz}\,\mathrm{s}^{-1}$) & $(-4.67 \pm 0.01) \times 10^{-15}$ & $-3.2^{+1.7}_{-1.6} \times 10^{-15}$ \\\hline
    \end{tabular}
    \caption{Glitch model parameters returned by the HMM (via \textsc{lmfit}) and \textsc{temponest} for the new small glitch described in Section \ref{subsec:new_gl_params}. In both cases the glitch epoch is taken to be immediately after the last pre-glitch ToA, to match the HMM. With \textsc{temponest} we do not fit for $t_\text{g}$, instead fitting for the unphysical parameter $\Delta\phi$. The given ranges are the 90\% confidence intervals in the \textsc{lmfit} case, and 90\% credible intervals in the \textsc{temponest} case.}
    \label{tbl:48636_gl_params}
\end{table}
The \textsc{tempo2} phase residuals before and after fitting for the glitch with \textsc{temponest} are shown in Fig. \ref{fig:tempo_residuals}.
Note that the post-fit residuals in Fig. \ref{fig:tempo_residuals} are not flat.
The \textsc{temponest} phase model includes timing noise as a sum of sinusoids \citep{LentatiAlexander2013}, and the sinusoids are not subtracted from the phase when forming residuals in \textsc{tempo2}.
Hence the sinusoidal structure in the post-fit residuals reflects the analytic form of the timing noise assumed in the \textsc{temponest} analysis.
The rms of the post-fit residuals is $2.5 \times 10^{-4}\,\mathrm{s}$.
To compare this value with the value expected from the \textsc{temponest} results, we assume that the residuals from \textsc{temponest} are dominated by the lowest-frequency component of the red noise process, which has frequency $T^{-1}$ where $T = 0.459\,\mathrm{yr}$ is the total timespan of the data section containing the glitch.
Given the returned red noise parameters $A = 10^{-8.8}\,\mathrm{yr}^{3/2}$, $\beta = 8.3$, we expect rms residuals $\sigma \propto \sqrt{P(T^{-1})}$ on the order of \begin{equation} \sigma =  \frac{1}{\sqrt{T}}\frac{A}{\sqrt{12}\pi}\left(\frac{T^{-1}}{f_{\text{yr}}}\right)^{-\beta/2}, \label{eqn:tn_rms_exp}\end{equation} i.e. $\sigma \approx 2.7 \times 10^{-4}\,\mathrm{s}$, in agreement with the post-fit rms residuals.
Equation \ref{eqn:tn_rms_exp} assumes an effective bandwidth $\sim 1/T$ to match the spacing of frequency components used in \textsc{temponest}.

With a fractional glitch size of $\Delta f/f = (8.19 \pm 0.04) \times 10^{-10}$ (according to the HMM estimate), this is the second smallest glitch reported in Vela.\footnote{The smallest is a glitch with $\Delta f/f = 3.9 \times 10^{-10}$ which was reported by \citet{JankowskiBailes2015}, although subsequently \citet{LowerBailes2020} have suggested that a timing noise-only model is favored.}

\section{Other small glitches}
\label{sec:other_small_gl}
In this work we do not re-analyse formally the several large ($\Delta f/f \gtrsim 10^{-6}$) glitches which occurred during the 24-year observing span, as their existence is almost certainly beyond dispute, and is verified by a quick and informal check with the HMM.
However, for the sake of completeness, and to compare with the results of previous work, we do investigate the two small glitches reported by \citet{EspinozaAntonopoulou2021}, using the HMM.
This section presents the findings of the investigation.
In short, both events exceed the HMM's Bayes factor threshold for detection, viz. $\ln K_\text{th} = 1.15$.

\subsection{A small glitch in 1991}
\label{subsec:1991_gl}
\citet{EspinozaAntonopoulou2021} reported a small glitch in Vela at MJD $48550.37(2)$ with $\Delta f = 6.21(3) \times 10^{-8}\,\mathrm{Hz}$.
We first check whether the HMM re-detects this glitch, analysing data between MJD $48475$ and MJD $48630$.
We find a single glitch candidate during the ToA gap between MJD $48549.9$ and MJD $48551.9$, with $\ln\left[K_1(k^*)\right] = 107.9$ ($k^* = 57$).
This is a clear detection.

We then estimate the glitch parameters taking the same approach as in Section \ref{subsec:new_gl_params}, using both the HMM and \textsc{temponest} separately.
The results of these analyses are shown in Table \ref{tbl:1991_gl_params}, along with the parameter estimates from \citet{EspinozaAntonopoulou2021}.
As the evolution of $f^*(t_n)$ is nonlinear over the full post-glitch timespan, we fit only the first 30 days' worth of post-glitch evolution to the HMM output, and only fit for $\Delta f_\text{p}$ and $\Delta\dot{f}_\text{p}$.
This approach matches the approach taken in \citet{EspinozaAntonopoulou2021}.
For the \textsc{temponest} fits, we do not truncate the data, but nor do we fit for a $\Delta\ddot{f}_\text{p}$ term.
\begin{table*}
    \centering
    \begin{tabular}{lrrr }\hline
        Parameter (units) & HMM fit (via \textsc{lmfit}) & \textsc{temponest} fit & \citet{EspinozaAntonopoulou2021} \\\hline
        $t_\text{g}$ (MJD) & $48549.9$ & $48549.9$ & $48550.37 \pm 0.03$\\
        $\Delta\phi$ (turns) & N/A & $(0.4 \pm 1.3) \times 10^{-3}$ & N/A \\
        $\Delta f_\text{p}$ (Hz) & $(5.86 \pm 0.01) \times 10^{-8}$ & $(5.7 \pm 0.1) \times 10^{-8}$ & $(6.21 \pm 0.05) \times 10^{-8}$\\
        $\Delta\dot{f}_\text{p}$ ($\mathrm{Hz}\,\mathrm{s}^{-1}$) & $(-5.6 \pm 0.1) \times 10^{-15}$ & $-0.3^{+2.3}_{-2.2} \times 10^{-15}$ & $(-16.5 \pm 0.3) \times 10^{-15}$ \\\hline
    \end{tabular}
    \caption{Glitch model parameters returned by the HMM (via \textsc{lmfit}) and \textsc{temponest} for the small glitch of 1991 described in Section \ref{subsec:1991_gl}.
    In the second and third columns the glitch epoch is taken to be immediately after the last pre-glitch ToA, to match the HMM. With \textsc{temponest} we fit for $\Delta\phi$ instead of $t_\text{g}$.
    The given ranges are the 90\% confidence intervals in the second and fourth columns, and 90\% credible intervals in the third column.}
    \label{tbl:1991_gl_params}
\end{table*}
\begin{figure*}
\centering
\includegraphics[width=0.45\linewidth]{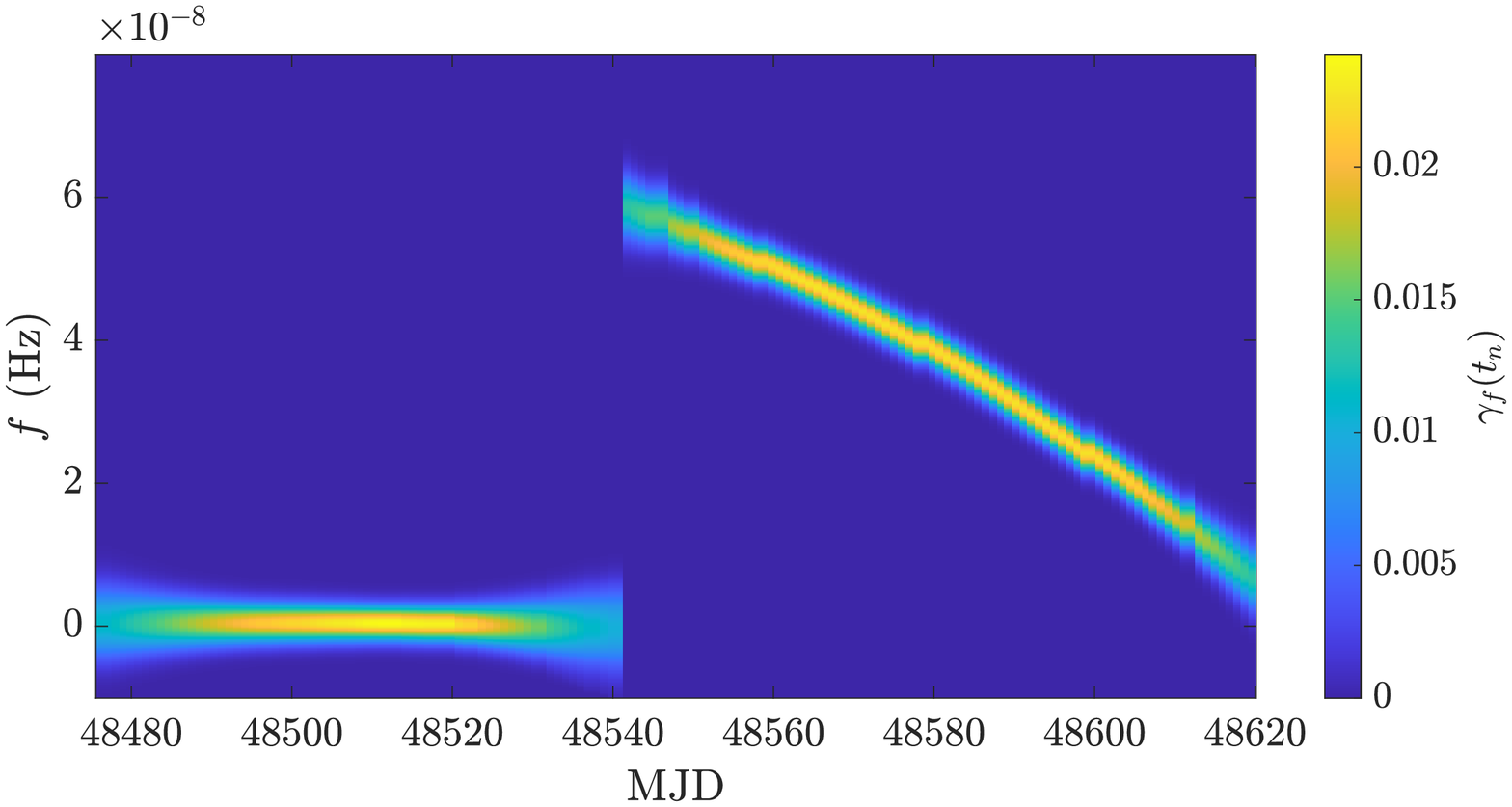} \includegraphics[width=0.45\linewidth]{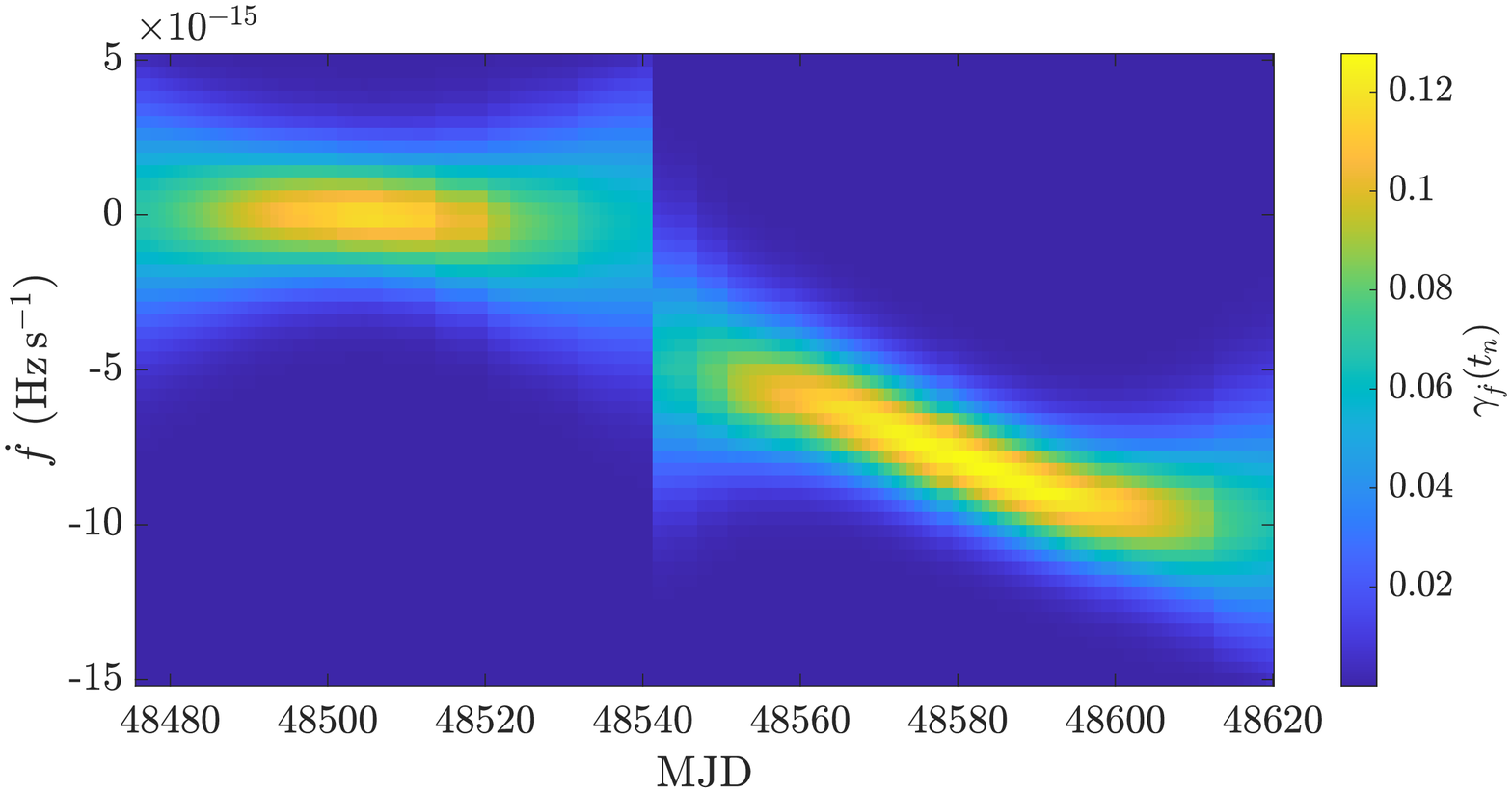}\\
\includegraphics[width=0.45\linewidth]{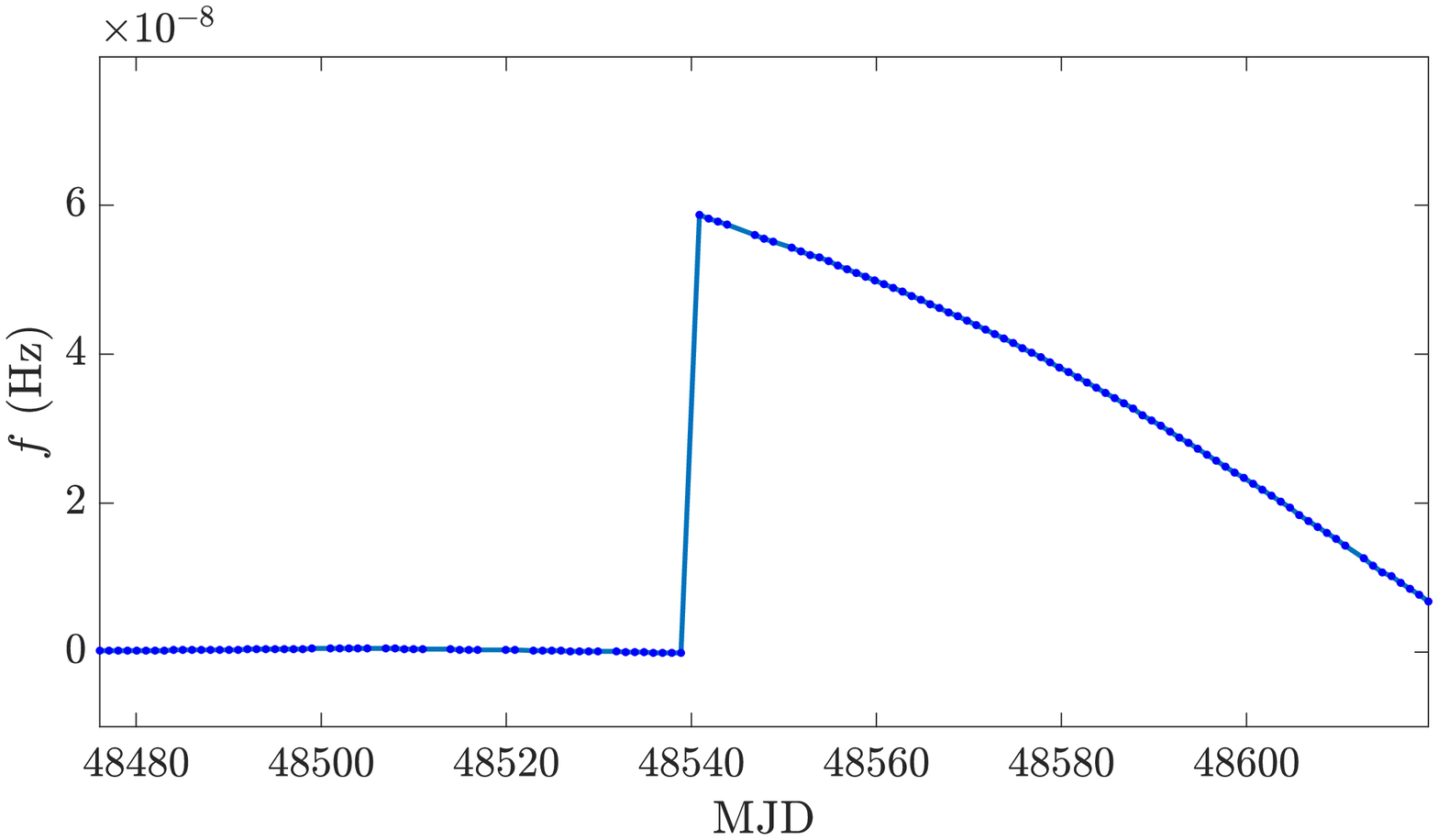} \includegraphics[width=0.45\linewidth]{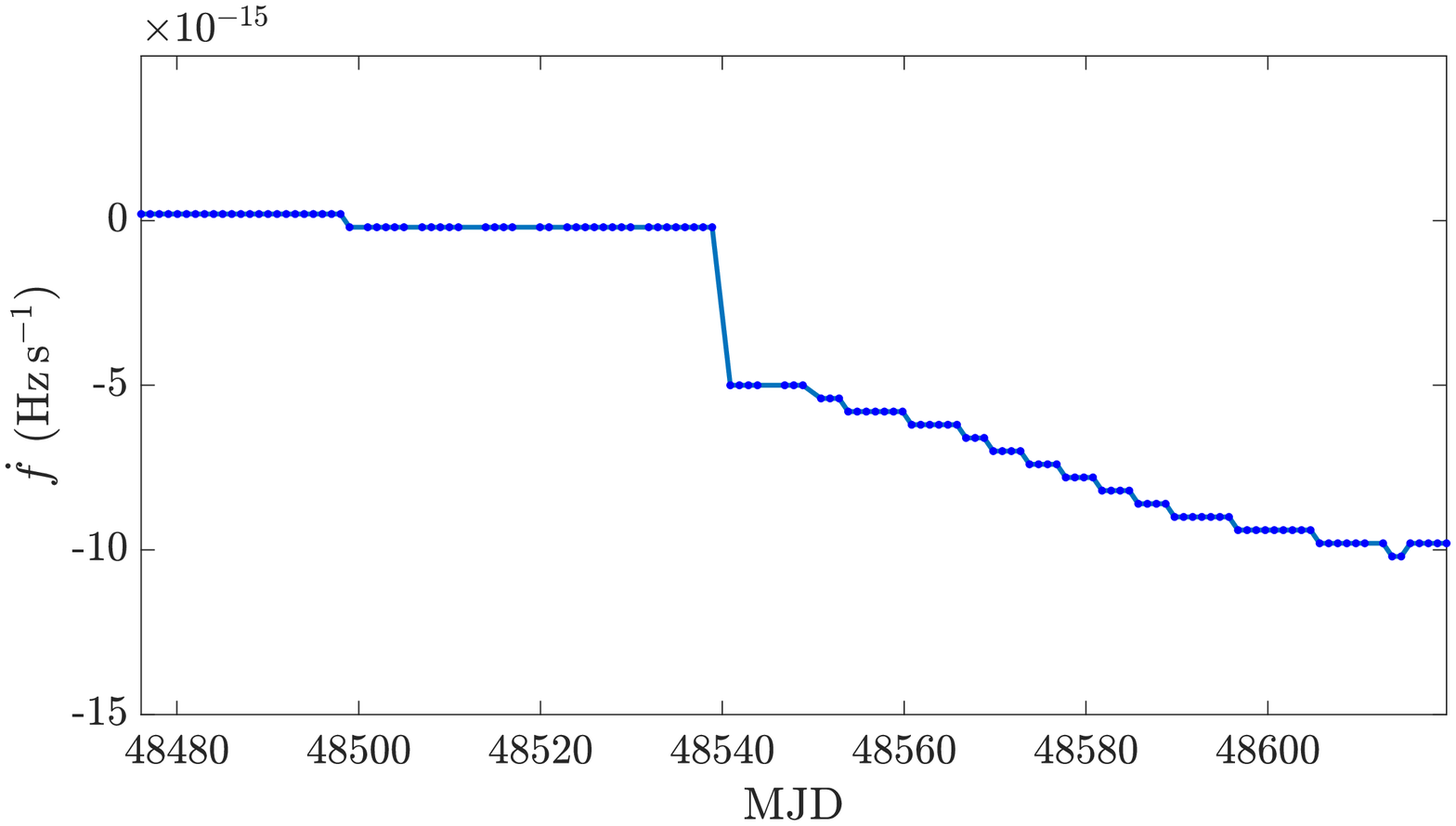}
\caption{\emph{(Top)} Heatmaps of the marginalised posterior distributions of $f$ \emph{(left)} and $\dot{f}$ \emph{(right)} for the data section containing the 1991 glitch described in Section \ref{subsec:1991_gl}. \emph{(Bottom)} Sequence of \emph{a posteriori} most probable $f$ \emph{(left)} and $\dot{f}$ \emph{(right)} states over time, for the same data section. The values of $f$ and $\dot{f}$ are relative to the pre-glitch timing solution, hence they are flat and centred on zero in the pre-glitch region.}
\label{fig:1991_gl_results}
\end{figure*}
The estimates of $\Delta f_\text{p}$ and $\Delta\dot{f}_\text{p}$ from the HMM, \textsc{temponest}, and \citet{EspinozaAntonopoulou2021} differ significantly.
All three 90\% credible (or confidence, for \textsc{temponest}) intervals are disjoint for both parameters.
This is to be expected.
The three methods take different approaches to measuring the glitch parameters and incorporating the effects of timing noise, and so the underlying models which the estimates are conditional upon differ substantially.
As an example of the effect that including timing noise can have on the recovered glitch parameters, fitting with \textsc{temponest} but excluding timing noise from the model produces a set of glitch parameters which is nearly disjoint from the other three sets listed in Table \ref{tbl:1991_gl_params}: $\Delta f_\text{p} = (6.68 \pm 0.05) \times 10^{-8}\,\mathrm{Hz}$ and $\Delta\dot{f}_\text{p} = (-2.4 \pm 0.8) \times 10^{-15}\,\mathrm{Hz}\,\mathrm{s}^{-1}$.

\subsection{A small glitch in 1999}
\label{subsec:1999_gl}
\citet{EspinozaAntonopoulou2021} also reported the detection of a small glitch in Vela at MJD $51425.12$, with a detectable exponential recovery.
Analysis with the HMM of the data between MJD $51294$ and MJD $51558$ reveals a glitch candidate during the ToA gap between MJD $51424.8$ and $51425.9$ with $\ln[K_1(k^*)] = 1.4 \times 10^{4}$ ($k^* = 75$).

\begin{figure*}
\centering
\includegraphics[width=0.45\linewidth]{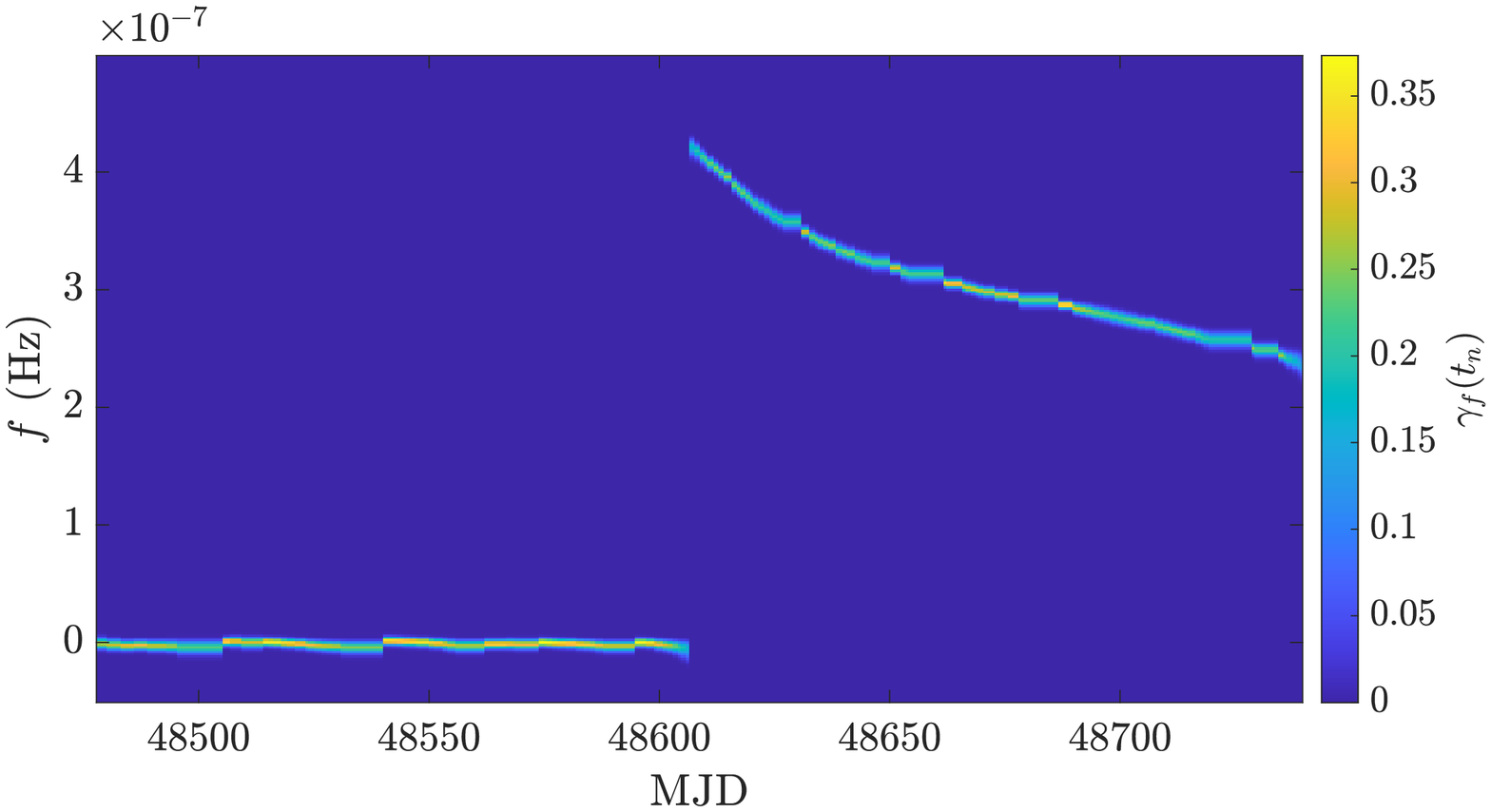} \includegraphics[width=0.45\linewidth]{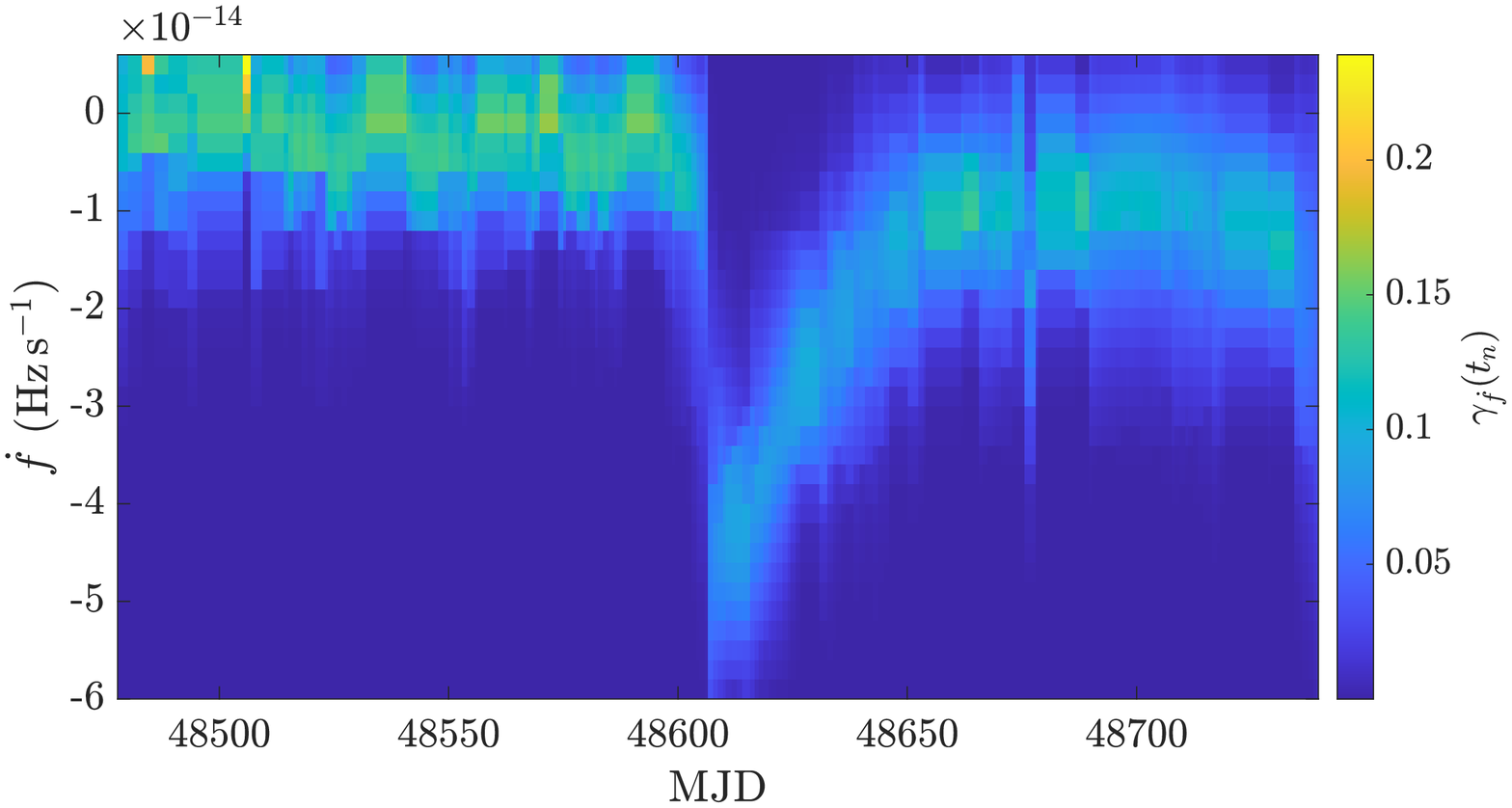}\\
\includegraphics[width=0.45\linewidth]{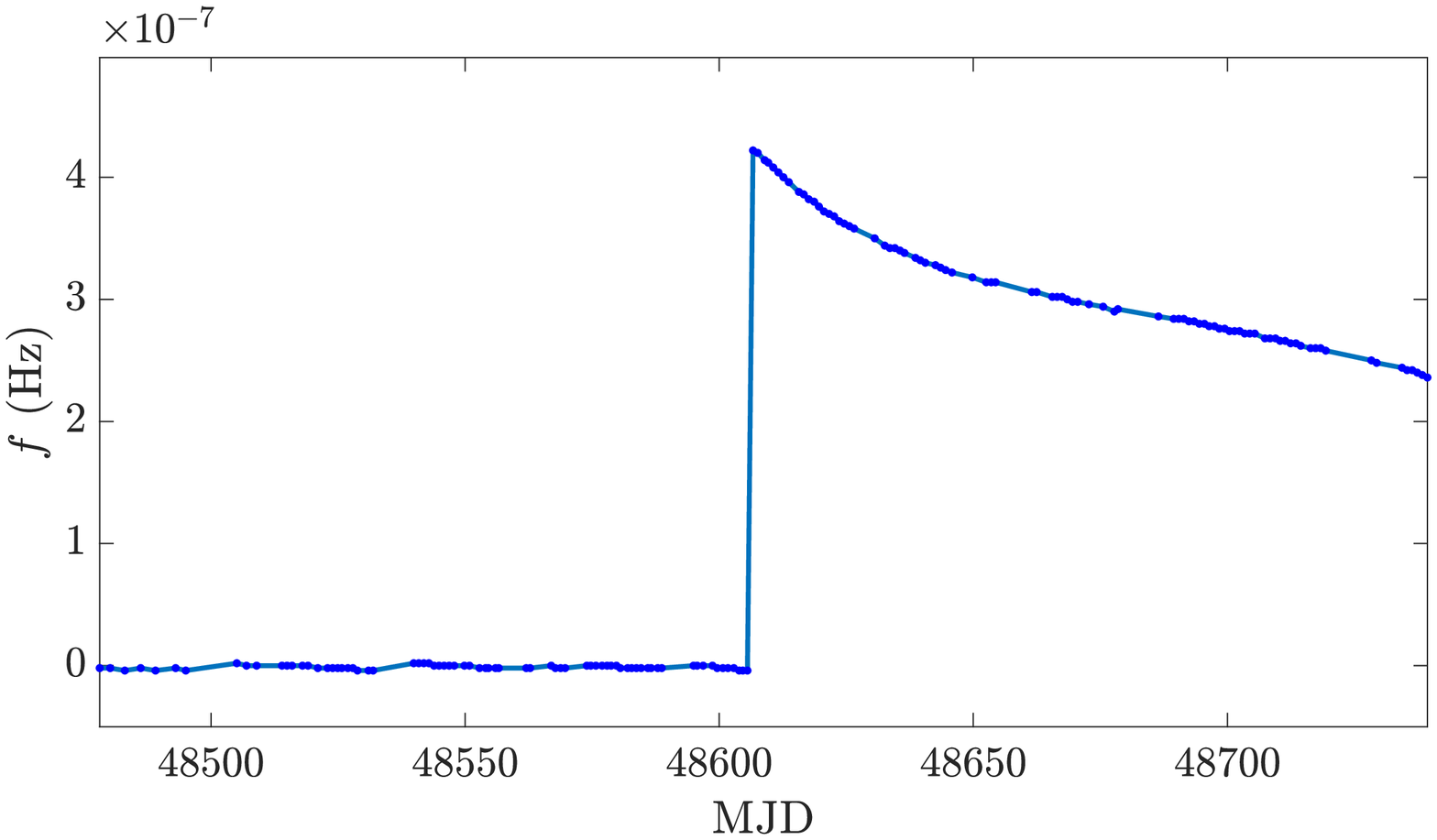} \includegraphics[width=0.45\linewidth]{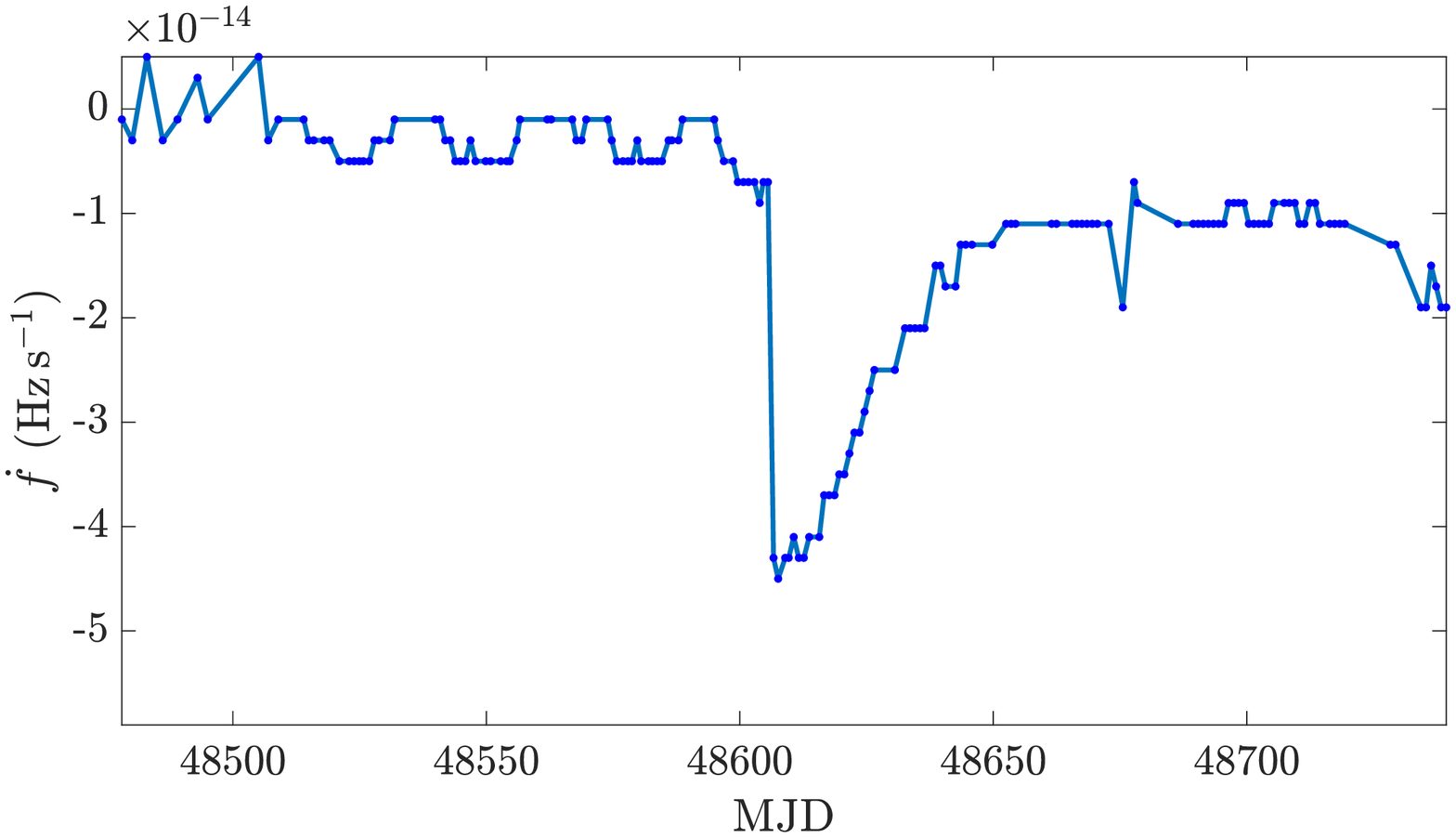}
\caption{\emph{(Top)} Heatmaps of the marginalised posterior distributions of $f$ \emph{(left)} and $\dot{f}$ \emph{(right)} for the data section containing the 1999 glitch described in Section \ref{subsec:1999_gl}. \emph{(Bottom)} Sequence of \emph{a posteriori} most probable $f$ \emph{(left)} and $\dot{f}$ \emph{(right)} states over time, for the same data section. The values of $f$ and $\dot{f}$ are relative to the pre-glitch timing solution, hence they are flat and centred on zero in the pre-glitch region.}
\label{fig:1999_gl_results}
\end{figure*}

As in Section \ref{subsec:1991_gl}, we calculate the posterior distribution of the hidden state and maximum \emph{a posteriori} tracks in $f$ and $\dot{f}$, the results of which are shown in Fig. \ref{fig:1991_gl_results}.
Interestingly the HMM finds the event successfully, even though exponential post-glitch recoveries are not part of the HMM's dynamical model \citep{MelatosDunn2020, DunnLower2021}.
This confirms the event's significance; an HMM model including post-glitch recoveries (a nontrivial extension which is the goal of future work) would produce an even higher Bayes factor, i.e. $\ln[K_1(k^*)] > 1.4 \times 10^4$.
To accommodate the exponential decay which is visible in Fig. \ref{fig:1999_gl_results} we increase artificially the value of the parameter which controls the strength of the timing noise included in the HMM, $\sigma_\text{TN}$, to $5 \times 10^{-17}\,\mathrm{Hz}\,\mathrm{s}^{-3/2}$ (see Section 3.4 of \citealt{MelatosDunn2020} for a detailed discussion of the timing noise model used here).
We then estimate the glitch parameters again using \textsc{lmfit} to fit the post-glitch evolution of $f^*(t_n)$ [this time including the exponential decay term in equation (\ref{eqn:gl_with_exp})] and separately using \textsc{temponest} to fit the full dataset.
The results are shown in Table \ref{tbl:1999_glitch_params}.
\renewcommand{\arraystretch}{1.3}
\begin{table*}
    \centering
    \begin{tabular}{lrrr}\hline
        Parameter (units) & HMM fit (via \textsc{lmfit}) & \textsc{temponest} fit & \citet{EspinozaAntonopoulou2021} \\\hline
        $t_\text{g}$ (MJD) & 51424.85 &  51424.85 & $51425.12 \pm 0.02$\\
        $\Delta\phi$ (turns) & N/A & $9.3^{+8.7}_{-9.4} \times 10^{-4}$ & N/A \\
        $\Delta f_\text{p}$ (Hz) & $(3.47 \pm 0.02) \times 10^{-7}$ & $2.7^{+0.9}_{-1.1} \times 10^{-7}$ & $(2.51 \pm 0.05) \times 10^{-7}$ \\
        $\Delta\dot{f}_\text{p}$ ($\mathrm{Hz}\,\mathrm{s}^{-1}$) & $(-0.91 \pm 0.03) \times 10^{-15}$ & $(0.2 \pm 1.7) \times 10^{-14}$ & $(0.8 \pm 0.2) \times 10^{-14}$\\
        $\Delta\ddot{f}_\text{p}$ ($\mathrm{Hz}\,\mathrm{s}^{-2}$) & N/A & N/A & $(-2.6 \pm 0.2) \pm 10^{-21}$ \\ 
        $\Delta f_\text{d}$ (Hz) & $(0.80 \pm 0.02) \times 10^{-7}$ & $1.6^{+1.0}_{-0.9} \times 10^{-7}$ & $(1.74 \pm 0.05) \times 10^{-7}$\\
        $\tau_\text{d}$ (days) & $17.5 \pm 1.1$ & $31.6^{+17.9}_{-10.6}$ & $31 \pm 2$\\\hline
    \end{tabular}
    \caption{Glitch model parameters returned by the HMM (via \textsc{lmfit}) and \textsc{temponest} for the small glitch of 1999 described in Section \ref{subsec:1999_gl}.
    In the second and third columns the glitch epoch is taken to be immediately after the last pre-glitch ToA, to match the HMM. With \textsc{temponest} we fit for $\Delta\phi$ instead of $t_\text{g}$. The given ranges are the 90\% confidence intervals in the second and fourth columns, and 90\% credible intervals in the third column.}
    \label{tbl:1999_glitch_params}
\end{table*}
\renewcommand{\arraystretch}{1.0}
As in Section \ref{subsec:1991_gl} we note that the estimates of the glitch parameters differ substantially between the three methods, as expected, because the estimates are conditional on three different signal models.
The HMM and \textsc{temponest} fits are not significantly improved by including a $\Delta\ddot{f}_p$ term, so we quote results from analyses which do not include this term.

\section{Undetected glitches}
\label{sec:undetected}
\subsection{Upper limits}
\label{subsec:ul}
In addition to searching for new glitches, we place upper limits on the magnitude of undetected glitches using synthetic data injections using \textsc{libstempo} \citep{Vallisneri2020}.
The injected glitches include only a permanent step in frequency, with no change in frequency derivative or exponential recovery included.
We quote $90\%$ frequentist upper limits for each section of data, denoted $\Delta f^{90\%}$.
Frequentist upper limits are based on the expected detection rate over many trials.
In particular the 90\% frequentist upper limits are the glitch sizes for which we expect to detect a glitch of that size 9 times out of 10, on average.
A detailed description of the procedure used to estimate $\Delta f^{90\%}$ values can be found in Appendix B of \citet{DunnMelatos2022}, except that we opt not to inject red timing noise into the synthetic datasets.
Experience indicates that although accounting for timing noise is an important part of setting up the HMM glitch detector, the difference between injecting timing noise and not is quite small when setting upper limits, as long as the HMM parameters match what has been used in the search. For example, when setting upper limits on a stretch of data between MJD 48551 and MJD 48719, we find that the value of $\Delta f^{90\%}$ is $1.6 \times 10^{-8}\,\mathrm{Hz}$ without timing noise injected, and $1.7 \times 10^{-8}\,\mathrm{Hz}$ with timing noise injected with parameters $A = 10^{-8.8}\,\mathrm{yr}^{3/2}$ and $\beta = 8.3$ [see equation (\ref{eqn:tn_rn_psd})].
\begin{figure*}
    \centering
    \includegraphics[width=0.9\textwidth]{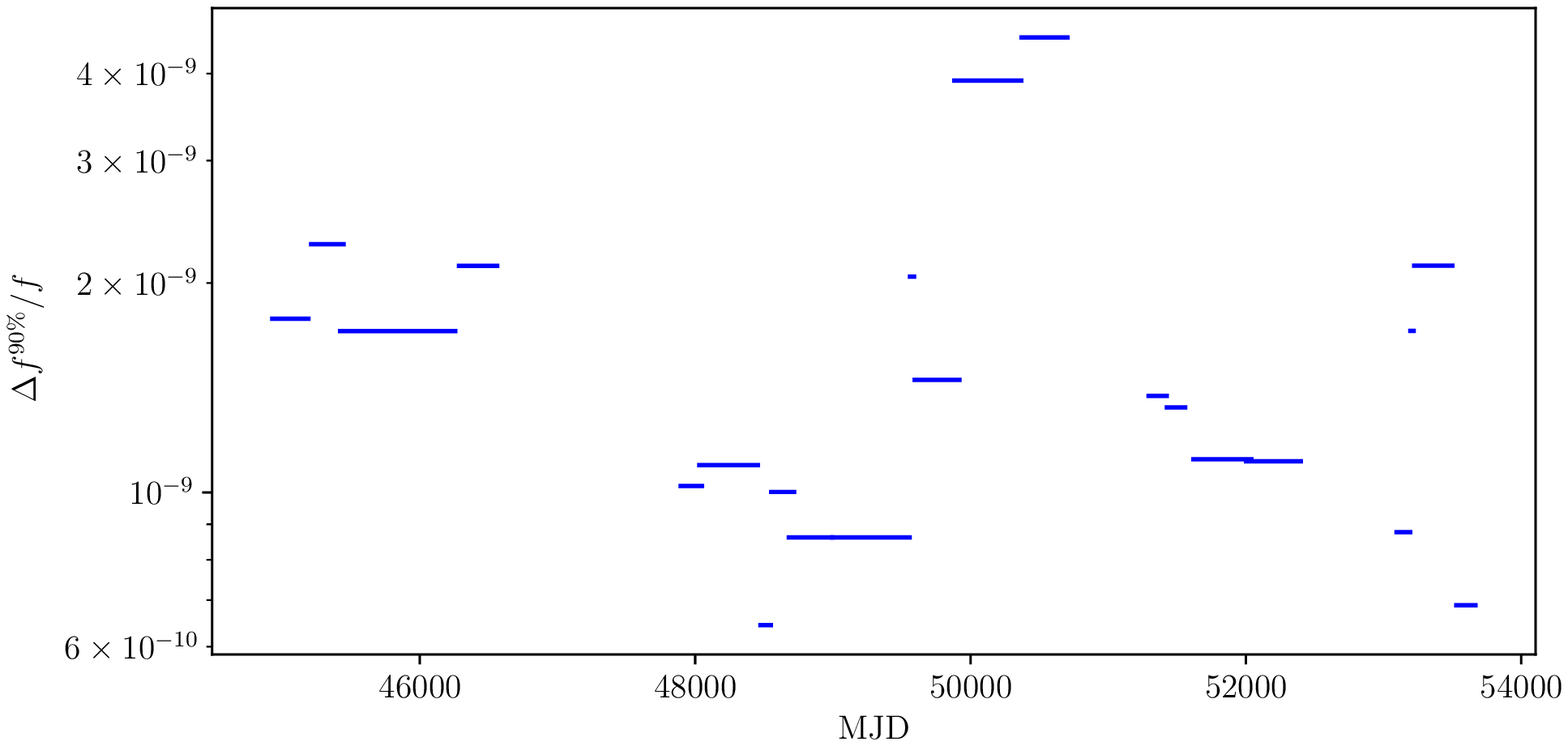}
    \caption{Frequentist HMM upper limits on the fractional size of undetected glitches $\Delta f^{90\%}/f$ across the 24 years of data analysed here.}
    \label{fig:upper_limits}
\end{figure*}

Fig. \ref{fig:upper_limits} shows the fractional upper limits $\Delta f^{90\%}/f$ obtained over the full dataset.
Values on the order of $10^{-9}$ are typical, with significant variation between sections of the dataset.
In the section of data containing the newly reported glitch the fractional upper limit is $1.0 \times 10^{-9}$, and the median fractional upper limit across the full dataset is $1.35 \times 10^{-9}$.
These limits are broadly consistent with the detection limits of \citet{EspinozaAntonopoulou2021}, which are roughly between $\Delta f/f = 10^{-10}$ and $\Delta f/f = 10^{-8}$, allowing for the dependence on $\Delta\dot{f}$ in those limits.
It is challenging to construct Fig. \ref{fig:upper_limits} without the automated and computationally fast algorithm provided by the HMM.

\subsection{Cadence of observations}
\label{subsec:cadence}
Somewhat counterintuitively, the upper limits derived from the HMM are elevated by the high cadence of the observations.
Across a single ToA gap of length $z$, the uncertainty in frequency is roughly \citep{MelatosDunn2020} \begin{equation} \delta f_1 = \left(\sigma_\text{ToA,1}^2 + \sigma_\text{ToA,2}^2\right)^{1/2}z^{-1},\end{equation} where $\sigma_\text{ToA,i}$ are the uncertainties on the two ToAs comprising the gap, measured in turns.
Evidently this quantity increases as $z$ decreases, which degrades the sensitivity of the HMM to small glitches.
However, we have the freedom to analyse only a subset of the data in order to increase $z$.
Sensitivity also degrades for long $z$: the timing noise included in the HMM means that jumps in frequency during long ToA gaps may be accomodated as stochastic wandering rather than detected as a glitch.
The scale of frequency wandering due to timing noise included in the HMM is given by \begin{equation} \delta f_2 = \sigma_\text{TN}z^{3/2}, \label{eqn:freq_wandering_tn}\end{equation} where $\sigma_\text{TN}$ is a free parameter which controls the strength of the timing noise, defined by equation (\ref{eqn:hmm_sigma_tn}).
In this work $\sigma_\text{TN}$ is on the order of $10^{-19}\,\mathrm{Hz}\,\mathrm{s}^{-3/2}$ to $10^{-18}\,\mathrm{Hz}\,\mathrm{s}^{-3/2}$.
In balancing the low- and high-$z$ behaviour, we seek to minimise the quadrature sum $\sqrt{\delta f_1^2 + \delta f_2^2}$ with respect to $z$.
This typically leads to an optimal cadence of $7$--$10$ days, significantly longer than the near-daily cadence of much of the data analysed here.
However, given the data sectioning described in Table \ref{tbl:sections}, reducing the data to achieve this optimal cadence frequently yields fewer than 15 ToAs per section, limiting the ability of the HMM to track the evolution of the pulsar.
Hence in this work we do not analyse subsets of the data to achieve the optimal cadence.

We emphasise that the considerations presented here are specific to the HMM method, rather than a generic discussion of the effect of observing cadence on glitch sensitivity.
\section{Conclusion}
In this paper we present the results of an HMM-based search for undetected glitches in the Vela pulsar using the high-cadence dataset from Mount Pleasant Observatory spanning 24 years.
We discover one new glitch, with fractional size $\Delta f/f = (8.19 \pm 0.04) \times 10^{-10}$, making it the smallest glitch in Vela which is clearly distinguished from timing noise.
We also confidently detect two other small glitches first reported by \citet{EspinozaAntonopoulou2021}, and re-measure the parameters of these glitches using both the HMM and \textsc{temponest}.
The parameter estimates are broadly consistent with one another and with the estimates given by \citet{EspinozaAntonopoulou2021}, when one considers that they are conditional on three different signal models.
We also employ Monte Carlo simulations to place 90\% frequentist upper limits on the size of undetected glitches, obtaining a median fractional upper limit of $\Delta f^{90\%}/f = 1.35 \times 10^{-9}$.
The latter exercise benefits from the automated and computationally fast nature of the HMM.

These results represent further steps towards a comprehensive understanding of glitch statistics.
Such an understanding is crucial to discriminating between falsifiable, microphysics-agnostic models of the glitch mechanism on the basis of long-term glitch statistics, e.g. auto- and cross-correlations between sizes and waiting times \citep{MelatosHowitt2018, CarlinMelatos2019}.
Quantifying the completeness of glitch catalogues is also important for analyses which are not directly aimed at investigating glitches, but are nonetheless sensitive to their presence, for example measurements of braking indices of glitching pulsars \citep{LynePritchard1996, EspinozaLyne2017, LowerJohnston2021}.
For pulsars of particular interest (e.g. PSR J0534$+$2200 or PSR J0537$-$6910) it may well be worth applying similar techniques as those employed here to search for missed glitches.

\section*{Acknowledgements}
Parts of this research are supported by the Australian Research Council (ARC) Centre of Excellence for Gravitational Wave Discovery (OzGrav) (project number CE170100004) and ARC Discovery Project DP170103625.
LD is supported by an Australian Government Research Training Program Scholarship.
CME acknowledges support from the ANID FONDECYT grant 1211964.
DA acknowledges support from an EPSRC/STFC fellowship (EP/T017325/1).
This work was performed in part on the OzSTAR national facility at Swinburne University of Technology. The OzSTAR program receives funding in part from the Astronomy National Collaborative Research Infrastructure Strategy allocation provided by the Australian Government.
This work was performed in part on the MASSIVE computing facility, with access provided through the National Computational Merit Allocation Scheme.
\section*{Data Availability}
The data underlying this article will be shared on reasonable request to the corresponding author.


\bibliographystyle{mnras}
\bibliography{vela_hmm}




\appendix
\section{MJD 48636 event as a candidate in previous searches}
\label{apdx:espinoza_48636}
The glitch event at MJD 48636 described in Section \ref{sec:small_glitch} has not been reported until now.
However, it did appear as a marginal candidate in the search for glitches presented by \citet{EspinozaAntonopoulou2021}.
In this appendix we briefly discuss the characteristics of this marginal candidate as it appeared in that search, and the reasons it was ultimately not recognised as a genuine glitch at that time.

After the two events which \citet{EspinozaAntonopoulou2021} reported as new glitches, the event at MJD 48636 was the next most significant.
It was ultimately discarded for the following reasons:
\begin{itemize}
    \item No $\Delta\dot{f}$ term was detected. This is due to a combination of factors. The automated glitch detection method fits only small stretches of data at once ($10$ or $20$ days), and the event at MJD 48636 is followed by just seven days of data and then a 2-week gap. These factors together are likely to be the reason that $\Delta\dot{f}$ was not detected.
    \item The detected $\Delta f$ value was relatively small, and comparable to many other similar events which were classified as noise. No quantitative difference was found between the event at MJD 48636 and these other events.
    \item In general large data gaps such as the one which follows a few days after this glitch were treated with caution, and strong claims about marginal candidates involving these large gaps were avoided.
\end{itemize}

\bsp	
\label{lastpage}
\end{document}